\documentclass[twoside,showpacs,twocolumn]{revtex4}
\usepackage{amsmath}
\usepackage{epsfig}
\usepackage{amsfonts}
\usepackage{amssymb}
\usepackage{graphicx}

\newcommand{\be}{\begin{equation}}
\newcommand{\ee}{\end{equation}}
\newcommand{\bea}{\begin{eqnarray}}
\newcommand{\eea}{\end{eqnarray}}

\begin{document}

\title{Neutron stars in the large-$N_{c}$ limit}
\author{Francesco Giacosa$^{(1,2)}$ and Giuseppe Pagliara$^{(3)}$}
\affiliation{$^{(1)}$~Institute of Physics, Jan Kochanowski University, ul.\
Swietokrzyska 15, 25-406 Kielce, Poland}
\affiliation{$^{(2)}$ Institute for Theoretical Physics, J. W. Goethe University,
Max-von-Laue-Str. 1, 60438 Frankfurt am Main, Germany}
\affiliation{$^{(3)}$Dip. di Fisica e Scienze della Terra dell'Universita' di Ferrara and
INFN Sez. di Ferrara, Via Saragat 1, I-44100 Ferrara, Italy}

\begin{abstract}
We study the phase transition from dense baryonic matter to dense quark
matter within the large-$N_{c}$ limit. By using simple constant speed of
sound equations of state for the two phases, we derive the scaling with $%
N_{c}$ of the critical quark chemical potential $\mu_{c}^{\mathrm{crit}}$
for this phase transition. While quark matter is strongly suppressed at
large $N_{c}$, the phase transition at a large but finite density could
nevertheless be important to determine the maximum mass of compact stars. In
particular, in the range $3\leq N_{c}\lesssim5.5$ the quark phase would take
place in compact stars and would lead to the formation of an unstable branch
of hybrid stars. As a consequence, the maximum mass is restricted to the
range $2.1M_{\odot}<M_{max}<3M_{\odot}$. For larger value of $N_{c}$, the
phase transition would occur at densities too high to be reached in the core
of compact stars. However, the very requirement that it occurs (although at
a very large density) translates into interesting constraints on the
stiffness of the baryonic phase: its speed of sound must exceed $\sqrt{1/3}$.
\end{abstract}

\maketitle

\section{Introduction}

The so-called large-$N_{c}$ method, where $N_{c}$ stands for the number of
color in Quantum\ Chromodynamics (QCD), is a useful theoretical approach to
investigate strong interactions in the nonperturbative regime of QCD \cite%
{thooft,witten,lebed,jenkins}. Namely, when $N_{c}$ is large, a series of
simplifications occurs: quark-antiquark mesons and glueballs retain their
masses but become very narrow, while baryon masses increase with $N_{c}$,
due to the fact that (at least) $N_{c}$ quarks are needed to form a white
state. The large-$N_{c}$ approach has been used to understand some
phenomenological features such as the OZI rule and it has been applied in
effective approaches of QCD in order to distinguish leading from sub-leading
terms.

More recently, the large-$N_{c}$ method has been also used to study some
features of the phase diagram of QCD \cite%
{ahrahrahr,kojo,sasakiredlich,bonanno,torrieri,lottini,heinznc}. Indeed it
represents an additional possible tool, often combined with effective
models, to describe (very) dense and hot matter (for a review of various
approaches, see for instance Ref. \cite{rischkerev}). In particular, in the
pioneering work of Ref. \cite{ahrahrahr}, McLerran and Pisarski make
interesting observations concerning the nature of a dense medium of hadrons:
a so-called quarkyonic phase, which is confined but chirally restored and
with a pressure proportional to $N_{c},$ is realized as an intermediate
phase between a `standard' baryonic phase at small density and a `standard'
quark-gluon phase at very high density. In a later study of the subject \cite%
{kojo}, the emergence of inhomogeneous condensations in the quarkyonic phase
was found to be favorable (this is in agreement with the recent results of
Refs. \cite{buballarev,carignano,wagner,heinz}).

The natural laboratory to test the properties of dense and strongly
interacting matter is the core of neutron stars. In these stellar objects
the central density can reach values up to ten times the nuclear matter
density and it is therefore conceivable that, besides nucleons, also heavier
baryons can take place, such as hyperons \cite%
{Weissenborn:2011ut,Chatterjee:2015pua} and delta resonances \cite%
{Drago:2014oja}, or that the phase transition to quark matter occurs. In
this respect, the well-established existence of neutron stars with masses of 
$2M_{\odot}$ (thus significantly larger than the canonical value of $%
1.4M_{\odot}$) offers a unique opportunity to test the stiffness of the
equation of state at high baryon densities. The question about the internal
composition of these massive stars is the subject of intense theoretical and
experimental/observational studies. At the moment, no firm conclusion for
instance can be drawn on whether hyperons or deltas do form in such systems.
While some calculations suggest that even in presence of those particles the
equation of state can be stiff enough to support $2M_{\odot}$ \cite%
{Weissenborn:2011ut,Maslov:2015msa}, other calculations find that the
threshold density for the appearance of hyperons is too large for those
particles to appear even in massive compact stars \cite{Lonardoni:2014bwa}.
Instead, other approaches find that the appearance of hyperons/delta soften
too much the equation of state and that massive objects should contain quark
matter, see for instance \cite{Chen:2015mda}. An alternative scenario, which
is supported by the indication of the existence of very compact objects
(which need to be confirmed by new observations, see \cite{Guillot:2013wu}),
is that two families of stars co-exist, baryonic stars and pure quark stars,
see Refs. \cite{Drago:2013fsa,Drago:2015cea,Drago:2015dea}. The possibility
of the occurrence of the phase transition to quark matter inside a compact
star (which would be then a hybrid star) is, similarly to the case of
hyperons and deltas, quite unsettled due to uncertainties which affect both
the baryonic matter equation of state and the quark matter equation of
state. Different possibilities and scenarios have been analyzed, see \cite%
{Bonanno:2011ch,Zdunik:2012dj,Kurkela:2014vha,Alford:2015dpa,Benic:2014jia}

Here we want to address the question about the phase transition to quark
matter in compact stars in the spirit of the large-$N_{c}$ limit and by
using the presently known constraints on the maximum mass of compact stars.
The advantage of the large $N_{c}$ approach is that the equation of state of
quark matter can be modeled by a simple prescription in which the speed of
sound is constant and equal to $1/\sqrt{3}$. Of course, we will need also to
establish the scaling with $N_{c}$ of the baryonic matter equation of state
to find the critical density for the transition. We will use also in this
case a prescription based on equations of state with constant speed of sound
in the regime of densities larger than about two times the nuclear
saturation energy density. At large $N_{c}$ the quark phase is clearly
suppressed, as found in \cite{ahrahrahr} but, as we will see, it could
nevertheless play an important role in determining the maximum mass of
neutron stars: a phase transition to quark matter occurring at large
densities would indeed lead to unstable hybrid stars configurations. The
critical density for the phase transition would correspond therefore to the
central density of the neutron star with the highest possible mass.

The paper is organized as follows: in Sec. II we discuss the quark phase,
the baryonic phase, and the first-order phase transition in the context of
the large-$N_{c}$ limit. In Sec. III we study the maximum mass of neutron
stars. Finally, in Sec. IV we present our conclusions.

\section{Equations of state at large $N_{c}$}

The study we are going to describe is performed by using, as customary, two
different models for the baryonic and the quark phase which are matched by
means of a Maxwell construction. In both cases we will present simple
parametrizations which however capture the main physical ingredients of the
(yet unknown) equation of state.

\subsection{Quark phase}

For modeling the equation of state of the quark phase we consider a free gas
of fermions with the additional contribution of a nonperturbative vacuum
pressure constant. Hence, the pressure $p_{q}$ reads: 
\begin{equation}
p_{q}=b_{1}N_{c}\mu_{q}^{4}-N_{c}^{2}B  \label{pq}
\end{equation}
where $b_{1}$, and $B$ are constants (independent on $N_{c}$ in the large-$%
N_{c}$ limit) and $\mu_{q}$ is the quark chemical potential. The first term
stands for the kinetic contribution of quarks:%
\begin{equation}
b_{1}=\frac{N_{f}}{12\pi^{2}}\text{ .}
\end{equation}
For three massless flavors one has $b_{1}=1/(4\pi^{2})$; this is the case
relevant for compact stars. We do not include the perturbative $\alpha_{s}$
corrections because they are vanishingly small in the large-$N_{c}$ limit.
The second term represents the non-perturbative contribution provided by the
vacuum pressure of QCD. This contribution is connected to the gluon
condensate and therefore takes a degeneracy factor of $N_{c}^{2}$ (see the
Appendix). In this way, along the temperature axis of the QCD phase diagram,
one obtains a critical temperature that is large-$N_{c}$ independent, in
agreement with basic expectations of QCD \cite{ahrahrahr,satzlectures}, with
models implementing a bag constant \cite{fuzzy,giacosabag}, as well as
lattice simulations \cite{panero}. Using the dilaton potential (presented in
the Appendix), one finds $\sqrt{3}B^{1/4}\sim200$ MeV (the factor $\sqrt{3}$
is due to the normalization adopted in Eq. (\ref{pq})). The value is
somewhat larger than the values usually adopted within the MIT bag model 
\cite{jaffe}. However, for what concerns the large-$N_{c}$ limit in the
context of the MIT bag model, care is needed. Namely, the MIT bag constant
must be $N_{c}$ independent in order to reproduce the correct scaling of the
masses of hadrons (see the detailed discussion in Refs. \cite%
{shuryak,zakharov}). For our purposes, we are interested in the
thermodynamic behavior of a gas of quarks (and gluons), hence the bag is
directly related to the nonperturbative QCD vacuum which scales as $N_{c}^{2}
$ (see detailed discussion in the Appendix).

It is easy to show that the equation of state (\ref{pq}) allows for the
existence of bound quark matter: the pressure indeed vanishes for 
\begin{equation}
p_{q}=0\leftrightarrow \mu _{q0}=\left( \frac{B}{b_{1}}N_{c}\right)
^{1/4}\propto N_{c}^{1/4}\text{ .}  \label{muq0}
\end{equation}%
By using the thermodynamics relation $p_{q}=\mu _{q}n_{q}-\varepsilon _{q}$,
where $n_{q}$ is the quark density and $\varepsilon _{q}$ the quark energy
density, one finds that the energy per baryon for this type of bound quark
matter $(e/n)_{q}\propto N_{c}^{5/4}$ whereas its baryon density $%
n_{b}=n_{q}/N_{c}\propto N_{c}\mu _{q0}^{3}/N_{c}=N_{c}^{3/4}$.

Notice that the energy per baryon of bound quark matter grows with $N_c$
faster than the baryon mass (which scales as $N_c$). This implies that the
so-called Witten hypothesis on the absolute stability of (strange) quark
matter \cite{Witten:1984rs} is not fulfilled in this limit. On the other
hand, as shown in \cite{bonanno}, in Walecka type models for nuclear matter,
the energy per baryon of bound nuclear matter scales as $N_{c}$ if the $%
\sigma \equiv f_{0}(500)$ meson, responsible for the binding of nuclear
matter, is interpreted as a quarkonium state. Instead, within the four-quark
assignment of the $\sigma \equiv f_{0}(500)$ meson \cite%
{pelaezrev,gallas,heinz} bound nuclear matter ceases to exist already for $%
N_{c}=4,$ as found in Ref. \cite{bonanno}. In this scenario, the only bound
state of strongly interacting matter would be realized in the quark phase.
Yet, the corresponding phase would be metastable. This is due to the fact
that, even in the absence of stable nuclear matter, a Fermi gas of nucleons
sets in at $\mu _{q}^{\text{Fermi gas}}\propto N_{c}^{0}$ (when nuclear
matter is realized, one has that $\mu _{q}^{\text{nuclear matter}}\leq \mu
_{q}^{\text{Fermi gas}}$). Hence, a gas of nucleons is favoured w.r.t. the
formation of bound quark matter.



Moreover, even when $f_{0}(500)$ is predominantly a four-quark state, there
is another possibility which needs to be further investigated in the future:
at some large values of $N_{c}$ a new type of (loosely) bound nuclear matter
takes place again thanks to pion exchange \cite{bonannoprivate}. Namely, in
the large-$N_{c}$ limit, the pion potential becomes a binding Coulomb
potential (the pion mass does not scale, but the nucleon's mass increases
linearly with $N_{c}$). In this scenario, which could not be found in\ Ref. 
\cite{Bonanno:2011ch} due to the employed mean-field approximation, nuclear
matter would not exist between $N_{c}=4$ up to a maximal value which needs
to be determined in a future work. Also in this case, the Witten's
hypothesis is not realized in the large-$N_{c}$ limit.

We thus conclude that the possibility of stable quark matter does not take
place at large-$N_{c}$ but still could be a specific feature of our
``small'' $N_c$ world. 

\subsection{Baryonic phase}

The equation of state of baryonic matter at high density is also quite
uncertain due to the intrinsic difficulties of solving the nuclear many body
problem. A widely used approach is based on relativistic mean field
Lagrangians (similar to the Walecka model) with parameters which are fixed
by using the experimental constraints on symmetric nuclear matter. An
updated parametrization of this class of models is the SFHo model of Ref. 
\cite{Steiner:2012rk} in which also recent constraints on the symmetry
energy are fulfilled. It is common to consider the results for the equation
of state as computed in these type of models to be reliable up to energy
densities not larger than about twice saturation density. Beyond that value
the equation of state is completely unknown. We use here a simple approach
that has been used in several papers, see \cite%
{Kalogera:1996ci,Lattimer:2010uk,Bedaque:2014sqa}: we adopt the equation of
state obtained within the SFHo model up to twice saturation energy density $%
2e_{0}$ and for larger densities we use a constant speed of sound equation
of state whose pressure is given as a function of the baryonic chemical
potential $\mu_{b}$ by the relation: 
\begin{equation}
p_{b}=a_{1}\mu_{b}^{\alpha}-K\text{ ,}
\end{equation}
where the constants $a_{1}$ and $K$ are fixed by matching this simple Ansatz
with the SFHo model at $2e_{0}$ (i.e. by requiring that at that value of the
energy density the pressure and the baryon density are continuous; note, $K$
turns out to be positive). The constants $a_{1}$ and $K$ scale in general
with $N_{c}$ and we need to fix such scaling behaviors in order to construct
the phase transition with quark matter at large $N_{c}$. In order to
determine $a_{1}\equiv a_{1}(N_{c}),$ let us consider baryons as interacting
by the exchange of conventional vector mesons with mass $m_{V}\propto
N_{c}^{0}$ and with the dimensionless coupling constant $g_{V}\propto\sqrt{%
N_{c}}$ [Note, the same conclusion would be reached by any quark-antiquark
mesonic exchange. We use vector mesons for definiteness and because they are
known to be important for the interaction among nucleons]. The vector
interaction implies that the propagator 
\begin{equation}
\frac{g_{V}^{2}}{m_{V}^{2}}\propto N_{c}
\end{equation}
enters into the expressions of pressure and energy density. The next step is
to notice that the constant $a_{1}$ must have dimension $4-\alpha,$ such
that the pressure $p_{b}$ has the dimension energy$^{4}.$ Hence, $%
a_{1}(N_{c})$ must be of the type%
\begin{equation}
a_{1}(N_{c})\propto\left( \frac{g_{V}^{2}}{m_{V}^{2}}\right) ^{\frac {%
\alpha-4}{2}}\propto N_{c}^{\frac{\alpha-4}{2}}\text{ .}
\end{equation}
Hence 
\begin{equation}
a_{1}(N_{c})=\tilde{a}_{1}N_{c}^{\frac{\alpha-4}{2}}
\end{equation}
where $\tilde{a}_{1}$ is a large-$N_{c}$ independent constant with dimension 
$4-\alpha$ and can be fixed by using the $N_{c}=3$ SFHo equation of state as
explained before. Similarly, to fix the scaling of $K$ we assume that the
quark chemical potential corresponding to the zero of the baryonic pressure
is $N_{c}$ independent \footnote{%
In turn, the baryonic chemical potential corresponding to the zero of the
baryonic pressure scales as $N_{c}$, in agreement with the results of \cite%
{bonanno} in the case of the standard quarkonium assignment for the sigma
meson.}. That fixes 
\begin{equation}
K=\tilde{K}N_{c}{}^{(3\alpha-4)/2}\text{ .}
\end{equation}
As a consequence, in terms of $N_{c}$ and $\mu_{q}^{\alpha}$ the baryonic
pressure reads: 
\begin{equation}
p_{b}=\tilde{a}_{1}N_{c}^{\frac{3\alpha-4}{2}}\mu_{q}^{\alpha}-\tilde{K}%
N_{c}{}^{\frac{3\alpha-4}{2}}
\end{equation}
The baryon density $n_{b}$ and the energy density $\varepsilon_{b}$ read: 
\begin{align}
n_{b} & =\frac{dp_{b}}{d\mu_{b}}=a_{1}\alpha\mu_{b}^{\alpha-1} \\
\varepsilon_{b} & =n_{b}\mu_{b}-p_{b}=a_{1}(\alpha-1)\mu_{b}^{\alpha}+K
\label{energydensity}
\end{align}
Thus, the constant speed of sound is: 
\begin{equation}
v_{b}=\sqrt{\frac{dp_{b}}{d\varepsilon_{b}}}=\frac{1}{\sqrt{\alpha-1}}.
\end{equation}
By imposing causality, $v_{b}<1$, one obtains a first constraint on $\alpha$%
: 
\begin{equation}
\alpha\geq2\text{ .}
\end{equation}
Note, the case $\alpha=1$ would correspond to the non-causal excluded volume
prescription \cite{satarov,satzlectures} for which the baryon density
saturates to a constant value when increasing the baryon chemical potential.
The stiffest equation of state corresponds, in agreement with causality, to $%
\alpha=2$. Quite interestingly, in this limit the pressure is proportional
to $N_{c}$: 
\begin{equation}
p_{b}=\tilde{a}_{1}N_{c}\mu_{q}^{2}-\tilde{K}N_{c}\text{ .}
\end{equation}
This result is in agreement with the large-$N_{c}$ equation of state of
nuclear matter found in Ref. \cite{bonanno}. In this sense, the
proportionality to $N_{c}$ would hold in a very large range of values for
the chemical potential $\mu_{q}$. This property is also compatible with the
quarkyonic phase introduced in Ref. \cite{ahrahrahr}, in which a confined,
but chirally restored phase with a pressure proportional to $N_{c}$ is
realized from intermediate up to high densities.

\subsection{Quark-hadron first-order phase transition}

We now turn to the phase transition from hadronic degrees of freedom to
quark degrees of freedom. The very request of the existence of this
transition sets a second constraint on the value of the parameter $\alpha$.
Namely, the Maxwell construction reads: 
\begin{equation}
p_{b}=\tilde{a}_{1}N_{c}^{\frac{3\alpha-4}{2}}\mu_{q}^{\alpha}-\tilde{K}%
N_{c}{}^{\frac{3\alpha-4}{2}}=b_{1}N_{c}\mu_{q}^{4}-N_{c}^{2}B=p_{q}\text{ .}
\label{pb=pq}
\end{equation}
This equation shows that at large but finite $N_{c}$ and in the limit of
large $\mu_{q}$ the quark phase is favored only if $\alpha<4.$ On the other
hand, larger values of $\alpha$ would imply that asymptotically the baryonic
phase is the favored phase (actually, as we will show in the following, one
cannot even find a physical solution of Eq. (15) if $\alpha\geq4$). Hence,
we consider the limiting case $\alpha=4$ as being excluded; in other terms,
we assume that a first order phase transition to quark matter does occur at
some large but finite density. Summarizing the constraints on $\alpha$ are: 
\begin{equation}
2\leq\alpha<4\text{ .}  \label{range}
\end{equation}

Let us now determine the critical chemical potential $\mu_{q}^{\text{crit}}$
for the phase transition to quark matter as obtained by the Maxwell
construction (i.e. by imposing that the pressures of the two phases are
equal at fixed quark chemical potential). It reads:

\begin{equation}
\mu_{q}^{\text{crit}}=\left( \frac{BN_{c}}{b_{1}}\right) ^{1/4}\left[ 1+...%
\right] \text{ for }2\leq\alpha\leq\frac{16}{7}  \label{muqcrit1}
\end{equation}
and%
\begin{equation}
\mu_{q}^{\text{crit}}=\left( \frac{\tilde{a}_{1}}{b_{1}}\right) ^{\frac {1}{%
4-\alpha}}N_{c}^{\frac{3\alpha-6}{2(4-\alpha)}}\text{ for }\frac{16}{7}%
<\alpha<4\text{.}  \label{mucrit2}
\end{equation}
Note, the limit $\alpha\rightarrow4^{-}$ implies $\mu_{q}^{\text{crit}%
}\rightarrow\infty,$ in agreement with Eq. (\ref{range}).

(i) For $\frac{16}{7}<\alpha<4$ the critical chemical potential grows as $%
\mu_{q}^{\text{crit}}$ $\propto N_{c}^{\beta>1/4},$ therefore it is indeed
possible to neglect the vacuum pressure term of the quark pressure in Eq. (%
\ref{pb=pq}).

(ii) For $2\leq\alpha\leq\frac{16}{7}$ the vacuum pressure term is
important, and the algebraic solution of $\mu_{q}^{\text{crit}}$ is more
complicated. However, it takes place just after $\mu_{q}$ exceeds $%
\mu_{q,0}\propto N_{c}^{1/4}$ from Eq. (\ref{muq0}) [dots in Eq. (\ref%
{muqcrit1}) refer to large-$N_{c}$ suppressed terms]. In fact, just after $%
\mu_{q,0}$ the quark pressure becomes positive and grows with $N_{c}^{2},$
while the baryonic pressures grows as $N_{c}^{\frac{7\alpha-8}{4}}.$ One can
easily see that the baryonic pressure grows slower than the quark pressure
as function of $N_{c}$ only if $\frac{7\alpha-8}{4}\leq2,$ hence for $%
\alpha\leq\frac{16}{7}$ \footnote{%
Note that for the quark phase to have positive pressure at the phase
transition point, $\mu_{q}^{\mathrm{crit}}$ must scale at least as $%
N_{c}^{1/4}$. In turn this implies that the term proportional to ${\tilde K}$
in Eq.(15) is always sub-leading (at large $N_{c}$) with respect to the term
proportional to $\mu_{q}^{\alpha}$ and thus it can be neglected.}. 

Finally, the pressure as function of the baryon density takes the form%
\begin{equation}
p_{b}=\lambda n_{b}^{\frac{\alpha}{\alpha-1}}\text{ with }\lambda=\frac {1}{%
a_{1}^{\frac{1}{\alpha-1}}\alpha^{\frac{\alpha}{\alpha-1}}}.
\end{equation}
Notice that the proportionality constant $\lambda$ decreases for increasing $%
a_{1}$ (at fixed baryon density $n_{b}$): this corresponds to the very
well-known fact that the opening of new degrees of freedom causes a decrease
of the pressure (namely, the exclusion principle).

\section{Phase transition in neutron stars}

We now study under which conditions the phase transition occurs in a neutron
star. First, we need to determine the structure of neutron stars, in
particular the relation between mass and central energy density, for
different values of $\alpha$. This is easily done by solving the
Tolman-Oppenheimer-Volkoff structure equation and the results are displayed
in Fig. 1. It is important to remark that in the structure equation only the
relation between pressure and energy density enters, therefore the maximum
mass (of neutron stars) does not scale with $N_{c}$. A dependence on $N_{c}$
emerges only in the case in which the maximum mass is actually determined by
the phase transition to quark matter as we will discuss in the following.

Fig. 1 shows that the maximum mass $M_{\mathrm{max}}$ increases by
decreasing the value of $\alpha$: from $M_{\max}\sim2.1M_{\odot}$ for $%
\alpha=3.5$ to $M_{\max}\sim3M_{\odot}$ for $\alpha=2$. The range of values 
\begin{equation}
2\leq\alpha\lesssim3.5  \label{rangegeneral}
\end{equation}
is compatible with the existence of stars with masses of $2M_{\odot}$. It is
interesting to notice that $\alpha=4$ (thus speed of sound $\sqrt{1/3}$)
would lead to a maximum mass smaller than the observational constraints, as
found in the analysis of \cite{Bedaque:2014sqa}. In particular, by taking
into account the uncertainties on the equation of state of baryonic matter
for densities below $2e_{0}$, the central value of the maximum mass is of
about $1.88M_{\odot}$ \cite{Bedaque:2014sqa}. We remind that in our large-$%
N_{c}$ scheme, the requirement that the phase transition to deconfined quark
matter does take place at a certain critical density translates into the
condition $\alpha<4$, which in turn allows to obtain masses larger than $%
2M_{\odot}$ (see the example with $\alpha=3.5$). Therefore, even if the
phase transition to deconfined quark matter does not take place in compact
stars, its occurrence (at a certain large density) makes the existence of
neutron stars as massive as $2M_{\odot}$ possible. This conclusion is quite
remarkable and basically model independent.

We study now whether the critical density for the phase transition can be
reached in the center of neutron stars. We notice first that the stiffer the
baryonic equation of state, the earlier the phase transition to quark
matter. On the other hand, as one can see from Fig. 1, the central density
of the maximum mass configuration decreases when reducing the value of $%
\alpha$. Whether the phase transition occurs or not depends on the relative
magnitude of these two opposite effects. Moreover, the critical density will
increase with $N_{c}$ because the larger $N_{c}$ the more unfavoured the
quark matter equation of state.

Let us fix first $N_{c}=3$. By solving equation (\ref{pb=pq}) and by using
equation (\ref{energydensity}) (we set $\sqrt{3}B^{1/4}=200$ MeV for the
present discussion, see Appendix), one can compute the critical energy
density of the baryonic phase corresponding to the onset of the phase
transition (i.e. the onset of the mixed phase). The full points labeled with 
$N_{c}=3$ correspond to such critical density: for $\alpha=3.5$ the phase
transition would occur within stars with masses above $1.9M_{\odot}$. On the
other hand for $\alpha=2$, only for masses above $2.1M_{\odot}$ the phase
transition takes place. These possible phase transitions occur always at a
quite large value of the density and the formation of quark matter (in
particular the mixed phase) makes the equation of state so soft (due to the
jump to a speed of sound of $\sqrt{1/3}$ in the pure quark phase) that only
unstable hybrid stars branches are obtained \cite{Pagliara:2007ph}, see also 
\cite{Alford:2013aca} (case A). The onset of the phase transition would
therefore correspond to the maximum mass configuration. In this respect,
even if quark matter does not form in stable stellar objects, its appearance
determines the maximum mass of baryonic stars. From the $N_{c}=3$ analysis,
one further reduces the range (\ref{rangegeneral}), namely $\alpha$ must be 
\textit{smaller} than about $\alpha_{\max}\simeq2.5$ (the blue line) in
order to explain the existence of $2M_{\odot}$ stars:%
\begin{equation}
2\leq\alpha\lesssim2.5\text{ for }N_{c}=3\text{ .}
\end{equation}
Moreover one should not observe stars with masses larger than about $%
2.1M_{\odot}$ (in the case in which the equation of state is the stiffest,
black line). Of course, this conclusion depends on the value of the vacuum
constant (the adopted value is $B^{1/4}\simeq200$ MeV, see Appendix A) and
the value of $N_{c}$, which is fixed to three in this analysis.

Clearly, a large $N_{c}$ analysis with $N_{c}=3$ is questionable (although
for some specific baryonic observables even $N_{c}=3$ is large) and the
simple equation of state adopted for quark matter should be regarded with
care. Indeed, it has been shown in \cite{Fraga:2013qra} that perturbative
corrections are responsible for a significant modification of the quark
matter equation of state with respect to the simple prescription here
adopted. One should therefore study the effect of increasing $N_{c}$, thus
making perturbative corrections smaller and smaller. In Fig. 1, we indicate
the values of $N_{c}$ for which (for each value of $\alpha$) the phase
transition would occur at the center of the maximum mass configuration (see
the filled dashed points). One can notice that for $\alpha=2$, this value is
close to $N_{c}\sim5.5$ and decreases to $N_{c}\sim3.4$ for $\alpha=3.5$.
Therefore a clear conclusion can be drawn: it is enough to fix a value of $%
N_{c}\geq5.5$ to rule out completely the appearance of deconfined quark
matter in compact stars.

As a last point, to study the effect of the adopted value of the vacuum
pressure constant, we have then set $N_{c}=4$ and we have determined the
critical mass $M_{crit}$ for the occurrence of the phase transition as a
function of $\sqrt{3}B^{1/4}$ in the cases $\alpha=2$ and $\alpha=2.5$, see
insert in Fig.1. Even for the smallest value of $\sqrt{3}B^{1/4}\sim165$MeV
(hence, in the most favored case for quark matter), the phase transition
would occur for masses larger than about $2.4M_{\odot}$, and thus safely
above $2M_{\odot}$. It is clear than that by just increasing $N_{c}$ from $3$
to $4$, the appearance of quark matter in compact stars becomes unlikely,
unless future measurements would find compact stars with masses above $%
2.4M_{\odot}$. Conversely, if the presence of quark matter (possibly in the
form of stable strange quark matter) will be proven via other astrophysical
measurements e.g. precise radii measurements \cite%
{Drago:2013fsa,Drago:2015cea}, gravitational waves measurements \cite%
{Bauswein:2015vxa} and gamma-ray-bursts observations \cite%
{Drago:2015dea,Drago:2015qwa,Pili:2016hqo,Li:2016khf} then the occurrence of
this transition in compact stars would be a particular phenomenon of our $%
N_{c}=3$ world, such as the existence of bound nuclear matter as remarked in 
\cite{bonanno}.

\begin{figure}[ptb]
\vskip 1cm 
\begin{centering}
\epsfig{file=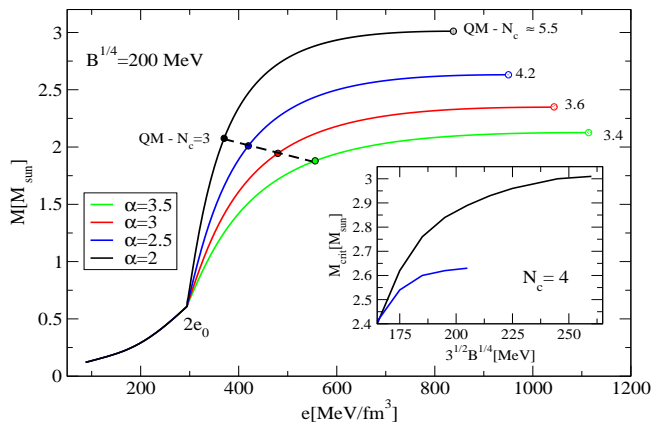,height=8.5cm,width=5.5cm,angle=-90}
\caption{Masses of neutron stars as functions of their central energy densities for different values
of $\alpha$. The filled points correspond to the critical mass for the transition to quark matter in the case $N_c=3$.
The filled dashed points correspond to the critical values of $N_c$ for which the phase transition
occurs in correspondence of the maximum mass configuration. In the insert we display the dependence of the critical mass
on the value of the vacuum pressure constant for two values of $\alpha$. }
\label{massa-raggio}
\end{centering}
\end{figure}

\section{Conclusions}

We have studied the phase transition from baryonic matter to deconfined
quark matter at high density by using equations of states that, both in the
quark- and in the baryonic phases, are motivated by the large-$N_{c}$
expansion in QCD. For the baryonic phase, we adopt (for energies above twice
nuclear matter saturation energy density) a constant speed of sound equation
of state in which the baryonic pressure is parametrized as follows $p_{b}
\propto\mu _{q}^{\alpha}$ where the parameter $\alpha$ determines the speed
of sound $v_{b}=\frac{1}{\sqrt(\alpha-1)}$. For the quark-phase, we have
used a free gas of quarks with the important inclusion of a gluon- (or
glueball-)driven vacuum pressure constant which scales as $N_{c}^{2}$
(details are explained in Appendix A).

By imposing causality ($v_{b}\leq1$) and by assuming that a first order
phase transition between baryonic matter and deconfined quark matter does
occur at a certain finite density, one obtains the following constraints on
the values of $\alpha$: $2\leq\alpha<4.$ In the large-$N_{c}$ limit, the
quark-hadron phase transition has the following scaling behavior: $\mu_{q}^{%
\text{crit}}\propto N_{c}^{1/4}$ for $2\leq\alpha<16/7$ and $\mu_{q}^{\text{%
crit}}\propto N_{c}^{\frac{3\alpha-6}{2(4-\alpha)}}$ for $16/7<\alpha<4.$
Hence, we continued the discussion introduced in\ Ref. \cite{ahrahrahr} at
large densities.

As an application of this simple formalism, we have investigated the effect
on the maximum mass of neutron stars of quark deconfinement. A first result,
which completes the analysis of \cite{Bedaque:2014sqa}, is that the
requirement of obtaining a phase transition to deconfined quark matter at
high density, $\alpha<4$, implies that the speed of sound of the baryonic
matter equation of state must exceed the value of $1/\sqrt{3}$. In turn,
this allows to fulfill the $2M_{\odot}$ limit. In this respect, the
occurrence of the phase transition at high density seems to be intimately
connected with the existence of massive neutron stars. A second result
concerns the values of $N_{c}$ for which the deconfined quark phase would
take place in astrophysical dense systems. While for $N_{c}=3$ (our world)
the phase transition limits the maximum mass to a value which is pretty
close to the observed $2M_{\odot}$ value, already at $N_{c}\simeq5.5$
deconfined quark matter would not play any role in the structure of compact
stars. In between, we have analyzed the case $N_{c}=4$: deconfinement of
quarks would start to be relevant for stars with masses $\geq2.4M_{\odot}$
(at least a candidate with a similar mass already exists \cite%
{vanKerkwijk:2010mt}). Finally, a promising possibility is to test the role
of quark matter in determining the maximum mass of compact stars via
gravitational waves observations: the threshold mass for prompt collapse in
binary neutron star mergers depends strongly on the value of the maximum
mass \cite{Bauswein:2013jpa,Bauswein:2017aur}. One could imagine that if the
maximum mass of neutron stars is set by the deconfinement of quarks (and
thus by a sudden softening of the equation of state) the temporal evolution
of the merger remnant could be qualitatively different with respect to the
standard scenario in which the phase transition is not considered.

\bigskip

\textbf{Acknowledgments}: The authors thank A. Drago, E. Maksymiuk, M.
Piotrowska, and A. Gazela-Zimolag for useful discussions. The work of F. G.
is supported by the Polish National Science Centre NCN through the OPUS
project no. 2015/17/B/ST2/01625.

\bigskip

\appendix

\section{The large-$N_{c}$ behavior of the vacuum pressure constant}

The Yang-Mills (YM) Lagrangian for an arbitrary number of colors $N_{c}$
reads (see, for instance, \cite{weisebook}): 
\begin{align}
\mathcal{L}_{YM} & =-\frac{1}{4}G_{\mu\nu}^{a}G^{a,\mu\nu}\text{ ,}  \notag
\\
G_{\mu\nu}^{a} &
=\partial_{\mu}A_{\nu}^{a}-\partial_{\nu}A_{\mu}^{a}+gf^{abc}A_{\mu}^{b}A_{%
\nu}^{c}\text{ ,}  \label{lqcd}
\end{align}
where $a=1,...,N_{c}^{2}-1$ and $f_{abc}$ are the structure constants of $%
SU(N_{c})$. For $N_{c}>1$, the YM Lagrangian contains 3-gluon and 4-gluon
vertices. The YM Lagrangian invariant under space-time dilatations, $x^{\mu
}\rightarrow x^{\prime\mu}=\lambda^{-1}x^{\mu}\,$, however this symmetry
does not survive quantization. The corresponding divergence reads: 
\begin{equation}
\partial_{\mu}J_{dil}^{\mu}=\frac{\beta(g)}{4g}G_{\mu\nu}^{a}G^{a,\mu\nu}%
\neq0\text{ , }\beta(g)=\mu\frac{\partial g}{\partial\mu}\text{ ,}
\label{sa}
\end{equation}
where the dimensionless coupling constant $g\equiv g(\mu)$ has become an
energy-dependent running coupling ($\mu$ is the energy scale at which the
coupling is probed). At the one-loop level, $\beta(g)=\mu\frac{\partial g}{%
\partial\mu}=-bg^{3}<0$ , $b=\frac{11N_{c}}{48\pi^{2}}$ , whose solution is $%
g^{2}(\mu)=\left( 2b\log\frac{\mu}{\Lambda_{YM}}\right) $ $,$where a
(Landau) pole at $\Lambda_{YM}$ is realized. Numerically, $%
\Lambda_{YM}\simeq250$ MeV: this number affects all hadronic processes. The
fact that $\beta(g)<0$ explains asymptotic freedom: the coupling $g(\mu)$
becomes smaller for increasing $\mu$. On the other side, for small $\mu$,
the coupling $g(\mu)$ increases. A (not yet analytically proven) consequence
is `confinement': gluons (and quarks) are confined in white hadrons. Notice
that $g$ scales as follows in the large-$N_{c}$ limit: $g\propto1/\sqrt{N_{c}%
}$ . This is the starting point of the study of the large-$N_{c}$ limit used
in this work.

A \textit{purely nonperturbative} consequence of the scale anomaly is the
emergence of a gluon condensate. Namely, the vacuum's expectation value of
the trace anomaly does not vanish:%
\begin{align}
\left\langle \partial_{\mu}J_{YM,dil}^{\mu}\right\rangle & =-\left\langle 
\frac{11N_{c}}{48}\frac{\alpha_{s}}{\pi}G_{\mu\nu}^{a}G^{a,\mu\nu
}\right\rangle  \notag \\
& \overset{N_{c}=3}{\sim}-\frac{33}{48}\left( 350\text{ MeV}\right) ^{4}%
\text{,}  \label{dilym}
\end{align}
where $\alpha_{s}=g^{2}/4\pi.$ The numerical results were obtained via
lattice, see Ref. \cite{sumrules,latticegc} and refs. therein. Note,the
vacuum's expectation value scales as 
\begin{equation}
-\left\langle \frac{11N_{c}}{48}\frac{\alpha_{s}}{\pi}G_{\mu\nu}^{a}G^{a,\mu%
\nu}\right\rangle \propto N_{c}^{2}\text{ . }
\end{equation}
Namely, $N_{c}\alpha_{s}$ is $N_{c}$-independent and the sum over $a$ goes
from $0$ to $N_{c}^{2}-1.$

Because of confinement, in the YM-vacuum glueballs are the relevant degrees
of freedom \cite{mainlattice}. The effective Lagrangian describing the trace
anomaly in terms of the ground-state scalar glueball $G$ reads \cite%
{migdal,salo}: 
\begin{align}
\mathcal{L}_{G}& =\frac{1}{2}(\partial _{\mu }G)^{2}-V_{dil}(G)\text{ , } 
\notag \\
V_{dil}(G)& =\frac{1}{4}\frac{m_{G}^{2}}{\Lambda _{G}^{2}}\left[ G^{4}\ln
\left( \frac{G}{\Lambda _{G}}\right) -\frac{G^{4}}{4}\right]   \label{dillag}
\end{align}%
By studying the fluctuations about the minimum, $G\rightarrow G_{0}+G$, one
can see that a field with mass $m_{G}$ emerges. This particle is the famous
scalar glueball. This is, according to lattice simulations \cite{mainlattice}%
, the lightest glueball with $m_{G}\sim 1.6$-$1.7$ GeV. The resonance $%
f_{0}(1710)$ is a very good candidate to describe the dilaton/glueball field 
$G$ \cite{stani,chenlattice,rebhan}. Note, the mass $m_{G}$ is independent
on $N_{c}$, while $\Lambda _{G}$ scales as $N_{c}$ in such a way that $G^{4}$%
-interaction scales as $1/N_{c}^{2}$ (this is the scaling of a four-leg
glueball term \cite{witten}):%
\begin{equation}
m_{G}\propto N_{c}^{0}\text{ , }\Lambda _{G}\propto N_{c}\text{ .}
\label{scaling}
\end{equation}%
The divergence of the dilatation Noether current of the dilaton field
presented in the Lagrangian (\ref{dillag}) is: 
\begin{equation}
\partial _{\mu }J_{dil,G}^{\mu }=G\partial _{G}V_{dil}(G)-4G=-\frac{1}{4}%
\frac{m_{G}^{2}}{\Lambda _{G}^{2}}G^{4}\text{ .}  \label{dilg}
\end{equation}%
By comparing Eqs. (\ref{dilym}) and (\ref{dilg}), one obtains ($N_{c}=3$)
(see also Ref. \cite{rattidrago}): 
\begin{equation}
\Lambda _{G}^{2}\simeq \frac{33}{12}\frac{(0.35\text{ GeV)}^{4}}{m_{G}^{2}}%
\simeq (0.12\text{ GeV)}^{2},
\end{equation}%
Finally, we notice that the YM vacuum energy reads: 
\begin{equation}
\varepsilon _{YM}=-N_{c}^{2}B=-\frac{m_{G}^{2}\Lambda _{G}^{2}}{16}\text{ .}
\label{eym}
\end{equation}%
This equation shows that the $\varepsilon _{YM}$ scales as $N_{c}^{2}$ (see
Eq. (\ref{scaling})), hence our assumption in Eq. (\ref{pq}) is justified.
Moreover, we obtain the following numerical values for $\sqrt{3}%
B^{1/4}\simeq 220$ MeV. When quarks are introduced, the constant $b$ changes
into $b=\frac{11N_{c}-2N_{f}}{48\pi ^{2}}.$ For $N_{f}=3$, a slight
reduction of $B$ is obtained: $\sqrt{3}B^{1/4}\simeq 214$ MeV. 

Strictly speaking, a negative vacuum's energy $\varepsilon _{YM}$ of Eq. (%
\ref{eym}) corresponds to a positive contribution $-\varepsilon _{YM}$ to
the dilaton/glueball, and hence to the hadronic, vacuum's pressure. Indeed,
the dilaton field with the potential in Eq. (\ref{dillag}) has been often
introduced in chiral hadronic models \cite%
{stani,Papazoglou:1996hf,Bonanno:2007kh,Bonanno:2008tt}. However, it is
convention to require that the hadronic matter pressure vanishes at zero
baryon chemical potential \cite{Papazoglou:1996hf}, hence one subtract this
contribution to the hadronic vacuum's pressure and, for consistency, to the
quark pressure as well. Summarizing, a negative contribution equal to $%
\varepsilon _{YM}=-N_{c}^{2}B$ appears in the expression for the pressure of the
quark phase, see Eq. (\ref{pq}).  Notice also that typically a chiral
dilaton hadronic model predicts the value of $B$ (which represents the
vacuum energy offset between the hadronic phase and the quark phase) would
be density dependent. This dependence shall not change the $N_{c}$ scaling,
but can change some quantitative features at finite density. We
retain this possibility for a future work.







\begin{thebibliography}{999}
\bibitem{thooft} 
G.~'t Hooft, 
Nucl.\ Phys.\ B \textbf{72} (1974) 461. 

\bibitem{witten} E.~Witten, 
Nucl.\ Phys.\ B \textbf{160} (1979) 57. 


\bibitem{lebed} 
R.~F.~Lebed, 
Czech.\ J.\ Phys.\ \textbf{49} (1999) 1273. 

\bibitem{jenkins} 
E.~E.~Jenkins, 
Ann.\ Rev.\ Nucl.\ Part.\ Sci.\ \textbf{48} (1998) 81. 

\bibitem{ahrahrahr} 
L.~McLerran and R.~D.~Pisarski, 
Nucl.\ Phys.\ A \textbf{796} (2007) 83. 

\bibitem{kojo} 
T.~Kojo, Y.~Hidaka, L.~McLerran and R.~D.~Pisarski, 
Nucl.\ Phys.\ A \textbf{843} (2010) 37. 

\bibitem{sasakiredlich} 
L.~McLerran, K.~Redlich and C.~Sasaki, 
Nucl.\ Phys.\ A \textbf{824} (2009) 86. 

\bibitem{bonanno} L.~Bonanno and F.~Giacosa, 
Nucl.\ Phys.\ A \textbf{859} (2011) 49. 


\bibitem{torrieri} 
G.~Torrieri and I.~Mishustin, 
Phys.\ Rev.\ C \textbf{82}, 055202 (2010). 


\bibitem{lottini} 
S.~Lottini and G.~Torrieri, 
Phys.\ Rev.\ Lett.\ \textbf{107} (2011) 152301. 
S.~Lottini and G.~Torrieri, 
Phys.\ Rev.\ C \textbf{88} (2013) 024912. 

\bibitem{heinznc} 
A.~Heinz, F.~Giacosa and D.~H.~Rischke, 
Phys.\ Rev.\ D \textbf{85} (2012) 056005. 

\bibitem{rischkerev} 
D.~H.~Rischke, 
Prog.\ Part.\ Nucl.\ Phys.\ \textbf{52} (2004) 197. 

\bibitem{buballarev} 
M.~Buballa and S.~Carignano, 
Prog.\ Part.\ Nucl.\ Phys.\ \textbf{81} (2015) 39. 

\bibitem{carignano} 
S.~Carignano, D.~Nickel and M.~Buballa, 
Phys.\ Rev.\ D \textbf{82} (2010) 054009. 

\bibitem{wagner} 
A.~Heinz, F.~Giacosa, M.~Wagner and D.~H.~Rischke, 
Phys.\ Rev.\ D \textbf{93} (2016) no.1, 014007. 

\bibitem{heinz} A.~Heinz, F.~Giacosa and D.~H.~Rischke, 
Nucl.\ Phys.\ A \textbf{933} (2015) 34. 

\bibitem{Chatterjee:2015pua} D.~Chatterjee and I.~Vidana, 
Eur.\ Phys.\ J.\ A \textbf{52} (2016) no.2, 29. 


\bibitem{Weissenborn:2011ut} S.~Weissenborn, D.~Chatterjee and
J.~Schaffner-Bielich, 
Phys.\ Rev.\ C \textbf{85} (2012) no.6, 065802 Erratum: [Phys.\ Rev.\ C 
\textbf{90} (2014) no.1, 019904]. 

\bibitem{Drago:2014oja} A.~Drago, A.~Lavagno, G.~Pagliara and D.~Pigato, 
Phys.\ Rev.\ C \textbf{90} (2014) no.6, 065809. 

\bibitem{Maslov:2015msa} K.~A.~Maslov, E.~E.~Kolomeitsev and
D.~N.~Voskresensky, 
Phys.\ Lett.\ B \textbf{748} (2015) 369. 

\bibitem{Lonardoni:2014bwa} D.~Lonardoni, A.~Lovato, S.~Gandolfi and
F.~Pederiva, 
Phys.\ Rev.\ Lett.\ \textbf{114} (2015) no.9, 092301.

\bibitem{Chen:2015mda} H.~Chen, J.-B.~Wei, M.~Baldo, G.~F.~Burgio and
H.-J.~Schulze, 
Phys.\ Rev.\ D \textbf{91} (2015) no.10, 105002. 


\bibitem{Guillot:2013wu} S.~Guillot, M.~Servillat, N.~A.~Webb and
R.~E.~Rutledge, 
Astrophys.\ J.\ \textbf{772} (2013) 7. 

\bibitem{Drago:2013fsa} A.~Drago, A.~Lavagno and G.~Pagliara, 
Phys.\ Rev.\ D \textbf{89} (2014) no.4, 043014. 


\bibitem{Drago:2015cea} A.~Drago, A.~Lavagno, G.~Pagliara and D.~Pigato, 
Eur.\ Phys.\ J.\ A \textbf{52} (2016) no.2, 40. 


\bibitem{Drago:2015dea} A.~Drago and G.~Pagliara, 
Eur.\ Phys.\ J.\ A \textbf{52} (2016) no.2, 41. 


\bibitem{Bonanno:2011ch} L.~Bonanno and A.~Sedrakian, 
Astron.\ Astrophys.\ \textbf{539} (2012) A16. 


\bibitem{Zdunik:2012dj} J.~L.~Zdunik and P.~Haensel, 
Astron.\ Astrophys.\ \textbf{551} (2013) A61. 

\bibitem{Kurkela:2014vha} A.~Kurkela, E.~S.~Fraga, J.~Schaffner-Bielich and
A.~Vuorinen, 
Astrophys.\ J.\ \textbf{789} (2014) 127. 

\bibitem{Alford:2015dpa} M.~G.~Alford, G.~F.~Burgio, S.~Han, G.~Taranto and
D.~Zappal\u{A}~, 
Phys.\ Rev.\ D \textbf{92} (2015) no.8, 083002. 

\bibitem{Benic:2014jia} S.~Benic, D.~Blaschke, D.~E.~Alvarez-Castillo,
T.~Fischer and S.~Typel, 
Astron.\ Astrophys.\ \textbf{577} (2015) A40. 

\bibitem{satzlectures} 
H.~Satz, 
Lect.\ Notes Phys.\ \textbf{841} (2012) 1. 



H.~Satz, 
Nucl.\ Phys.\ A \textbf{862-863} (2011) 4. 

\bibitem{fuzzy} 
R.~D.~Pisarski, 
Prog.\ Theor.\ Phys.\ Suppl.\ \textbf{168} (2007) 276. 

\bibitem{giacosabag} 
F.~Giacosa, 
Phys.\ Rev.\ D \textbf{83} (2011) 114002. 

\bibitem{panero} 
M.~Panero, 
Phys.\ Rev.\ Lett.\ \textbf{103} (2009) 232001. 
B.~Lucini and M.~Panero, 
Phys.\ Rept.\ \textbf{526} (2013) 93. 

\bibitem{jaffe} E.~Farhi and R.~L.~Jaffe, 
Phys.\ Rev.\ D \textbf{30} (1984) 2379. 

\bibitem{shuryak} 
E.~V.~Shuryak, 
World Sci.\ Lect.\ Notes Phys.\ \textbf{71} (2004) 1 [World Sci.\ Lect.\
Notes Phys.\ \textbf{8} (1988) 1]. 

\bibitem{zakharov} 
W.~A.~Bardeen and V.~I.~Zakharov, 
Phys.\ Lett.\ \textbf{91B} (1980) 111. 


\bibitem{Witten:1984rs} E.~Witten, 
Phys.\ Rev.\ D \textbf{30} (1984) 272. 

\bibitem{pelaezrev} 
J.~R.~Pelaez, 
Phys.\ Rept.\ \textbf{658} (2016) 1.

\bibitem{gallas} 
S.~Gallas, F.~Giacosa and G.~Pagliara, 
Nucl.\ Phys.\ A \textbf{872} (2011) 13.

\bibitem{bonannoprivate} L. Bonanno, S. Lottini and A. Heinz, private
communications (2016).

\bibitem{Steiner:2012rk} A.~W.~Steiner, M.~Hempel and T.~Fischer, 
Astrophys.\ J.\ \textbf{774} (2013) 17. 

\bibitem{Kalogera:1996ci} V.~Kalogera and G.~Baym, 
Astrophys.\ J.\ \textbf{470} (1996) L61. 

\bibitem{Lattimer:2010uk} J.~M.~Lattimer and M.~Prakash, 
arXiv: 1012.3208. 

\bibitem{Bedaque:2014sqa} P.~Bedaque and A.~W.~Steiner, 
Phys.\ Rev.\ Lett.\ \textbf{114} (2015) no.3, 031103. 


\bibitem{satarov} L.~M.~Satarov, M.~N.~Dmitriev and I.~N.~Mishustin, 
Phys.\ Atom.\ Nucl.\ \textbf{72} (2009) 1390. 


\bibitem{Pagliara:2007ph}  G.~Pagliara and J.~Schaffner-Bielich,  
Phys.\ Rev.\ D \textbf{77} (2008) 063004.  

\bibitem{Alford:2013aca} M.~G.~Alford, S.~Han and M.~Prakash, 
Phys.\ Rev.\ D \textbf{88} (2013) no.8, 083013. 

\bibitem{Fraga:2013qra} E.~S.~Fraga, A.~Kurkela and A.~Vuorinen, 
Astrophys.\ J.\ \textbf{781} (2014) no.2, L25. 


\bibitem{Bauswein:2015vxa} A.~Bauswein, N.~Stergioulas and H.~T.~Janka, 
Eur.\ Phys.\ J.\ A \textbf{52} (2016) no.3, 56. 


\bibitem{Drago:2015qwa} A.~Drago, A.~Lavagno, B.~Metzger and G.~Pagliara, 
Phys.\ Rev.\ D \textbf{93} (2016) no.10, 103001. 


\bibitem{Pili:2016hqo} A.~G.~Pili, N.~Bucciantini, A.~Drago, G.~Pagliara and
L.~Del Zanna, 
Mon.\ Not.\ Roy.\ Astron.\ Soc.\ \textbf{462} (2016) no.1, L26. 


\bibitem{Li:2016khf} A.~Li, B.~Zhang, N.~B.~Zhang, H.~Gao, B.~Qi and T.~Liu, 
Phys.\ Rev.\ D \textbf{94} (2016) 083010. 


\bibitem{vanKerkwijk:2010mt} M.~H.~van Kerkwijk, R.~Breton and
S.~R.~Kulkarni, 
Astrophys.\ J.\ \textbf{728} (2011) 95. 


\bibitem{Bauswein:2013jpa} A.~Bauswein, T.~W.~Baumgarte and H.-T.~Janka, 
Phys.\ Rev.\ Lett.\ \textbf{111} (2013) no.13, 131101. 


\bibitem{Bauswein:2017aur} A.~Bauswein and N.~Stergioulas, 
arXiv:1702.02567 [astro-ph.HE]. 













\bibitem{weisebook} 
A.~W.~Thomas and W.~Weise, \textquotedblleft The Structure of the
Nucleon,\textquotedblright\ Berlin, Germany: Wiley-VCH (2001) 389 p.

\bibitem{sumrules} 
M.~A.~Shifman, A.~I.~Vainshtein and V.~I.~Zakharov, 
Nucl.\ Phys.\ B \textbf{147}, 385 (1979); 
L.~J.~Reinders, H.~R.~Rubinstein and S.~Yazaki, 
Nucl.\ Phys.\ B \textbf{186}, 109 (1981); 
J.~Marrow, J.~Parker and G.~Shaw, 
Z.\ Phys.\ C \textbf{37}, 103 (1987); 
B.~V.~Geshkenbein, 
Sov.\ J.\ Nucl.\ Phys.\ \textbf{51}, 719 (1990) [Yad.\ Fiz.\ \textbf{51},
1121 (1990)]; 
D.~J.~Broadhurst, P.~A.~Baikov, V.~A.~Ilyin, J.~Fleischer, O.~V.~Tarasov and
V.~A.~Smirnov, 
Phys.\ Lett.\ B \textbf{329}, 103 (1994); 
F.~J.~Yndurain, 
Phys.\ Rept.\ \textbf{320}, 287 (1999); 
B.~L.~Ioffe and K.~N.~Zyablyuk, 
Eur.\ Phys.\ J.\ C \textbf{27}, 229 (2003); 
K.~Zyablyuk, 
JHEP \textbf{0301}, 081 (2003) [arXiv:hep-ph/0210103]; 
A.~Samsonov, 
arXiv:hep-ph/0407199. 

\bibitem{latticegc} 
J.~Kripfganz, 
Phys.\ Lett.\ B \textbf{101}, 169 (1981); 
A.~Di Giacomo and G.~C.~Rossi, 
Phys.\ Lett.\ B \textbf{100}, 481 (1981); 
A.~Di Giacomo and G.~Paffuti, 
Phys.\ Lett.\ B \textbf{108}, 327 (1982); 
E.~M.~Ilgenfritz and M.~Muller-Preussker, 
Phys.\ Lett.\ B \textbf{119}, 395 (1982); 
S.~s.~Xue, 
Phys.\ Lett.\ B \textbf{191}, 147 (1987); 
M.~Campostrini, A.~Di Giacomo and Y.~Gunduc, 
Phys.\ Lett.\ B \textbf{225}, 393 (1989); 
A.~Di Giacomo, H.~Panagopoulos and E.~Vicari, 
Nucl.\ Phys.\ B \textbf{338}, 294 (1990); 
X.~D.~Ji, 
arXiv:hep-ph/9506413; 
G.~Boyd and D.~E.~Miller, 
arXiv:hep-ph/9608482. 

\bibitem{mainlattice} 
Y.~Chen, A.~Alexandru, S.~J.~Dong, T.~Draper, I.~Horvath, F.~X.~Lee,
K.~F.~Liu and N.~Mathur \textit{et al.}, 
Phys.\ Rev.\ D \textbf{73, }014516 (2006). 

\bibitem{migdal} 
A.~A.~Migdal and M.~A.~Shifman, 
Phys.\ Lett.\ B \textbf{114}, 445 (1982).

\bibitem{salo} 
C.~Rosenzweig, A.~Salomone and J.~Schechter, 
Phys.\ Rev.\ D \textbf{24}, 2545 (1981); 
A.~Salomone, J.~Schechter and T.~Tudron, 
Phys.\ Rev.\ D \textbf{23}, 1143 (1981); 
C.~Rosenzweig, A.~Salomone and J.~Schechter, 
Nucl.\ Phys.\ B \textbf{206}, 12 (1982) [Erratum-ibid.\ B \textbf{207}, 546
(1982)]; H.~Gomm and J.~Schechter, 
Phys.\ Lett.\ B \textbf{158}, 449 (1985); 
R.~Gomm, P.~Jain, R.~Johnson and J.~Schechter, 
Phys.\ Rev.\ D \textbf{33}, 801 (1986). 

\bibitem{stani} 
S.~Janowski, F.~Giacosa and D.~H.~Rischke, 
Phys.\ Rev.\ D \textbf{90} (2014) no.11, 114005; 
D.~Parganlija, P.~Kovacs, G.~Wolf, F.~Giacosa and D.~H.~Rischke, 
Phys.\ Rev.\ D \textbf{87} (2013) no.1, 014011

\bibitem{chenlattice} 
L.~-C.~Gui, Y.~Chen, G.~Li, C.~Liu, Y.~-B.~Liu, J.~-P.~Ma, Y.~-B.~Yang and
J.~-B.~Zhang, 
Phys.\ Rev.\ Lett.\ \textbf{110} (2013) 021601. 

\bibitem{rebhan} 
F.~Br\"{u}nner, D.~Parganlija and A.~Rebhan, 
Phys.\ Rev.\ D \textbf{91} (2015) no.10, 106002 Erratum: [Phys.\ Rev.\ D 
\textbf{93} (2016) no.10, 109903]. 

\bibitem{rattidrago} 
A.~Drago, M.~Gibilisco and C.~Ratti, 
Nucl.\ Phys.\ A \textbf{742} (2004) 165. 

\bibitem{Bonanno:2007kh}  L.~Bonanno, A.~Drago and A.~Lavagno,  
Phys.\ Rev.\ Lett.\ \textbf{99} (2007) 242301.  


\bibitem{Bonanno:2008tt}  L.~Bonanno and A.~Drago,  
Phys.\ Rev.\ C \textbf{79} (2009) 045801.  


\bibitem{Papazoglou:1996hf}  P.~Papazoglou, J.~Schaffner, S.~Schramm,
D.~Zschiesche, H.~Stoecker and W.~Greiner,  
Phys.\ Rev.\ C \textbf{55} (1997) 1499.  
\end{thebibliography}
\end{document}